# An Overview of Trust Standards for Communication Networks and Future Digital World

Huilin Wang, Xin Kang, *Senior Member, IEEE,* Tieyan Li, *Member, IEEE,* Zhongding Lei, *Senior Member, IEEE,* Cheng-Kang Chu, *Member, IEEE,* and Haiguang Wang, *Senior Member IEEE*

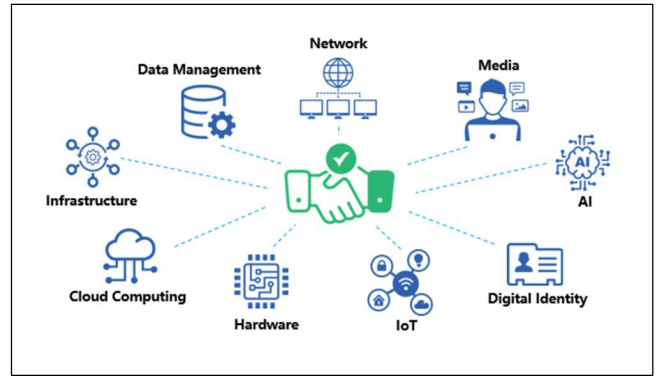

**Figure 1:** Active Areas of Trust Standards

*Abstract*—With the development of Information and Communication Technologies, trust has been applied more and more in various scenarios. At the same time, different organizations have published a series of trust frameworks to support the implementation of trust. There are also academic paper discussing about these trust standards, however, most of them only focus on a specific application. Unlike existing works, this paper provides an overview of all current available trust standards related to communication networks and future digital world from several main organizations. To be specific, this paper summarizes and organizes all these trust standards into three layers: trust foundation, trust elements, and trust applications. We then analysis these trust standards and discuss their contribution in a systematic way. We discusse the motivations behind each current in forced standards, analyzes their frameworks and solutions, and presents their role and impact on communication works and future digital world. Finally, we give our suggestions on the trust work that needs to be standardized in future.

## I. INTRODUCTION

TRUST has played an important role throughout the history of modern computer science development. Trust is defined as the degree of willingness of a party to be vulnerable and take a risk in the interaction with another based on specific expectation [1]. In the early period of computer security, the discussion of trust was more focusing on whether human should trust a program resilience from Trojan horse [2], while the discussion of trust is generalized to extensive fields from different aspects nowadays. Inspired from the way of trust establishment between a human and another in reality, scholars explore the trust relationship between human and objects (entities), objects and objects, and further to entities and entities in the digital world. The trustworthiness of an object or entity has also been a popularly pondered. Trustworthiness is defined as the capability whether a party has in order to satisfy another's trust or be relied by others [3]. Common methods to identify the trustworthiness of a party include but not limited to authentication and evaluation. Trust modelling is then proposed beyond trust relationships and trustworthiness with the idea of trust quantification. In trust modelling, trust becomes a scalar value representing trustee's trustworthiness in a trust relationship between trustor and trustee, and this result is produced by quantitative measurements of different influencing factors in the trust relationship. Trust and its applications have been extensively studied in academia over the past decades. Ting et al. reviewed conceptions of trust and surveyed comprehensive digital trust modelling techniques [4]. Wang et al. proposed a novel trust framework called SIX-Trust, which involves 3 layers: sustainable trust (S-Trust), infrastructure trust (I-Trust) and xenogenesis trust (X-Trust), to construct trustworthy and secure 6G networks [5]. However, there are few literatures studying about trust related standards.

Today, while trust has been widely implemented in security design rationales, organisations such as ITU, NIST, ISO and IETF have published a series of standards to supervise and regulate the application of trust in various fields of security. As shown in Fig1., these standards mainly distribute in seven areas of security: infrastructure, data management, network, media, AI, digital identity, IoT, hardware and cloud computing. However, there is no available literature giving a survey and overview of the existing trust standards on communication networks and future digital world, which motivates us to start this paper. The objective of this paper is to present a comprehensive overview of all current available trust standards related to communication networks from these main standard organizations. Specifically, this paper summarizes and organizes all these trust standards into three layers: trust foundation, trust elements, and trust applications. We then analysis these trust standards and discuss their contribution in a systematic way. To be specific, this paper discusses the motivations behind each current in forced standards, analyzes their frameworks and solutions, and presents their role and impact on communication works and future digital world.

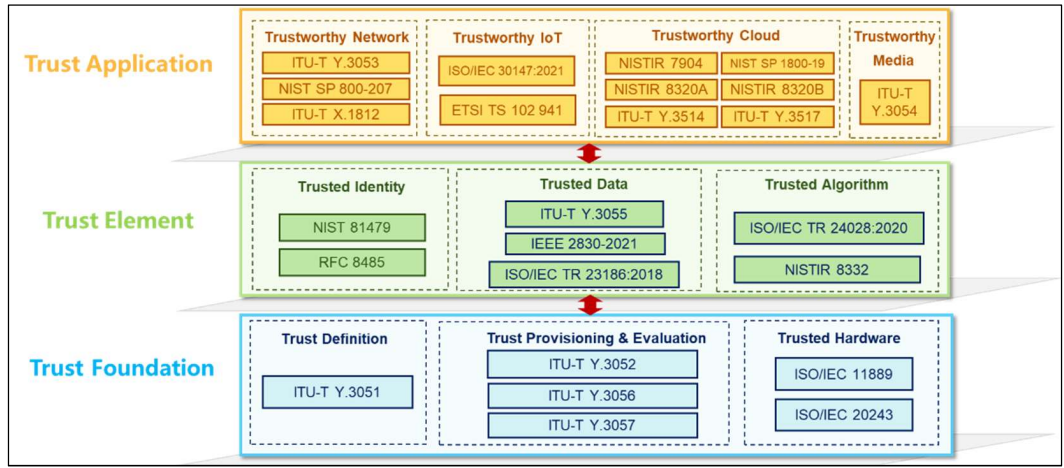

**Figure 2:** Overview of Trust Standards for Communication Networks and Future Digital World

## II. Trust Foundation

As shown in Fig.2, we summarize and categorize all the existing trust-related standards into three layers: trust foundation, trust element, and trust application. The standards in the trust foundation level serves as the basis for upper layers, since they focus on the fundamental concepts, definitions, evaluation of trust. In this section, we introduce the standards covering these fundamental things for trust.

### A. Trust Definitions and Trust Environment

Trust definitions and the related trusted environment play an indispensable role in development of ICT. The establishment of trust between entities requires interoperability and information security provided by trusted environment in ICT infrastructure, so that entities could diminish the risk and haze as much as possible by using trust to predicting the results of interaction.

The formal definition of trusted environment that is widely used today is defined in ITU-T Y.3051 in 2017 by ITU-T SG13, aiming to offer the anticipated level of confidence and protection to entities. This recommendation defines trusted environment as an environment that provides a set of technical and regulatory conditions to allows establishing trust between interacting agents within such environment. It highlights a series of requirements to build such trusted environment. This includes concerns on predictability, information security, interoperability and availability of administration services. The recommendation also outlines basic principles of trusted environment on both technical aspect and legal aspect, which refines the conception of trusted environment. This recommendation provides thorough conception of trusted environment in ICT infrastructure and service for further implementation of trust in different scenarios.

### B. Trust Provisioning and Trust Evaluation

On top of the trusted environment defined in ITU-T Y.3051, ITU-T Y.3052 proposed a trust framework for trust provisioning in ICT infrastructures and services, intending to resolve security issues in ICT caused by lack of trust. Trust is categorized to direct trust and indirect trust according to the conception of trust. The obligation of trust is introduced based on the analysis of risk in several circumstances in ICT, together with the elaboration of the concept and the fundamental characteristic of trust in the context of trusted ICT infrastructures and services. It then describes models for trust provisioning including social trust, cyber trust as well as physical trust. This recommendation also provides a trust evaluation framework as well as a detailed trust provisioning process on top of these models and the conception of trust.

ITU-T Y.3056 instead concentrates on future distributed ecosystems which involve entities requiring open access to trusted service as well as mutual identification, authentication and authorization. Such requirement could be satisfied by the security capabilities of devices and underlying network as well as standardization of related inferences and processes in ICT infrastructures. Therefore, by considering the security capability of network operators who take charge of connecting users and devices to Internet, ITU SG 13 proposes the framework of bootstrapping devices and applications in the ecosystem by network operators. Such framework allows network operators to share their network security capabilities with users and service or equipment providers to achieve open and secure access interactions in the ecosystem. Besides, the recommendation also provides a reference model and a functional architecture beyond the requirements to illustrate the elements, functions, reference points and security parameters of provisioning of the bootstrapping capabilities. The information flow is provided at the end to demonstrate the operation of bootstrapping processes.

According to the trust provisioning model provided in ITU-T Y.3052, the information of trust evaluation involves trust attribute, trust indicator and trust index. ITU SG 13 extended the conception of trust evaluation and proposed a trust index model for ICT infrastructure and services in ITU-T Y.3057 to provide an approach for trust evaluation that covers different characteristics of trust. Trust index is an overall accumulation of trust indicators, reflecting the evaluation and measurement of the trust degrees of entities. ITU SG 13 also defined a set of trust indicators based on characteristics of trust and fundamental criteria of trust evaluation. These trust indicators are categorized into objective trust indicators and subjective trust indicators to cover both objective trust and subjective trust.

*C. Trusted Hardware*

Root of trust is the start of chain-of-trust in system which all security and reliability of high-level functions, features and operations relies on. Since the root of trust is considered absolutely trusted, a common approach is to implement in hardware, because hardware is considered immune from malware attack due to its inalterability [7].

Trusted Platform Module (TPM) is a system component that could improve the security of the platform and realize trusted computing by establishing trust. TPM-based roots of trust in hardware solution could overcome the limitation of software-based solutions in resisting malware. The regulations of TPM, including the architecture, data structures, command interface and behavior, are defined in ISO/IEC 11889 by TCG, aiming to define the interaction between the host and the TPM. In the trusted platform, TCG defines a mechanism of establishing trust by identifying hardware and software components on the platform to ensure the trustworthiness of trusted platform and the service it provides. Such mechanism requires TPMs to provide three types of root of trust (RoT) under the hardware protection: measurement, storage and reporting, to describe characteristics that impact a platform's trustworthiness with minimum necessary functionality. Root of Trust for Measurement (RTM) is designed to reveal what software runs on a platform in a trusted manner. Root of Trust for Storage (RTS) mainly involves creating, managing, and keeping encryption keys and other data values. Root of Trust for Reporting (RTR) helps external entities establish trust in platform software measurements or encryption keys with the proof of the presence of a value in the TPM. These three types of RoT are realized by components of TPM. On top of the mechanism of RoT, TCG provides a generic library of commands, cryptographic algorithms and capabilities of TPMs in the rest of standards for flexible implementation purpose and to meet global various requirements in different deployment scenarios.

The factors that impact trustworthiness in hardware could be either own vulnerability or lack of robust hardware support [7]. These two problems could be chased back to supply chain where products are initially designed and produced. The main threats of products in supply chain are maliciously tainted product and counterfeit product. Maliciously tainted products may have backdoors that allows adversary to perform attack, and the integrity of counterfeiting products could not be verified. Therefore, The Open Group proposes Open Trusted Technology Provider Standard (O-TTPS) in ISO/IEC 20243 to mitigate the risks caused by tainted and counterfeit products. A set of guidelines, requirements and recommendations for suppliers and providers is addressed to resolve problems from tainting and counterfeiting which may threats to the integrity of Commercial Off-the-Shelf (COTS) ICT products throughout the product life cycle.

III. TRUST ELEMENT

The trust element layer summarizes the standards related to the three elements for future digital world and communication network. From our perspective, all kinds of activities in future communication networks and digital world can be regarded as a form of digital "transaction" in a broad sense, and a digital transaction is composed of three elements: identity, data, and algorithm. As shown in Fig.3, a digital transaction is realized by designed algorithm, takes in and produces data in the process, and happens between different entities with identities. The successful completion of a digital transaction relies on trust and trust relationship among different components and parties. Hence, trusted identity, trusted data and trusted algorithm are of great importance to ensure the trustworthiness of future communication networks and digital transaction.

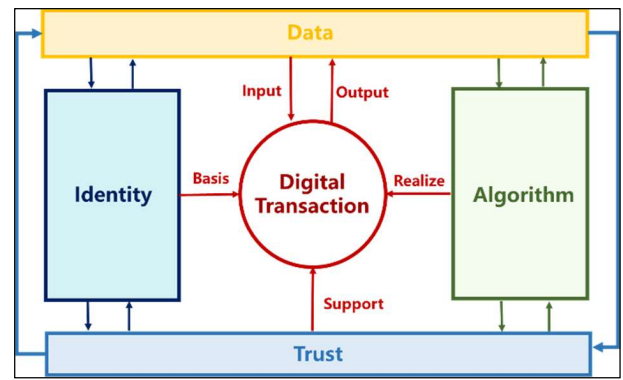

**Figure 3:** Illustration of a Digital Transaction

*A. Trusted Identity*

Identity is the key to the digital world. Every entity needs to be verified before being granted access to the digital world. However, identity management is very complicated and imposed heavy burden on the current digital system. Federated identity centralizes user trust to federated identity provider, and relies on the single-use token provided by trusted identity provider when granting right to user to access their services. Such identity ecosystem simplifies user access and mitigates threats of identity leaking as it only requires user to register their personal identifiable information at identity provider [8]. However, trust between entities in the identity ecosystem is still vital, and different service providers may have different risk management procedures, making risks of identity federation hard to identify and manage. Thus, NISTIR 8149 introduces the idea of trust framework to overcome the difficulties of risk management across multiple entities and support the establishment of mutual trust among them in identity federations. The trust framework consists four components: system rules, legal structure, establishing conformance and recognizing conformance. System rules specify the technical requirements, security requirement and required identity management operations in identity federations, while legal structure ensures the members in federations are glued legally. Establishing conformance provides assessments and methodologies for members in federations to evaluate their conformances, and recognizing conformance describes several mechanisms including registry or listing service, trust marks and digital certificates, for communication, conformation recognition and trust establishment in federation.

Another standard closely related to digital identity is IETF RFC 8485. RFC focuses on the measurement of trust in digital transaction. Basically, there are two approaches for the measurement of trust in digital identity transaction: one is to combine all indicators to a single scalar value, and the other is to evaluate the detailed set of attribute data locally to make trust decisions. For the first approach, the information trust values present is limited because trust attributes are compressed to a single scalar value, which makes the trust value incomparable in some cases. The second approach is limited by the requirement for identity provider (IdP) capable to collect store, and transmit the attributes data, and the requirement for rely parties (RP) to process data. Therefore, RFC presents the Vector of Trust (VoT) framework in RFC 8485 to simplify processing data for RP while being more expressiveness than single scalar value. A VoT contains four orthogonal components: identity proofing, primary credential usage, primary credential management, and assertion presentation. Any of the component could appear multiple times in a single vector to reflect changes over time, but a specific value of a vector component cannot appear multiple times in a single vector. The sample applications and metrics of VoT are also provided in the document.

*B. Trusted Data*

Data especially personal data has become the key and gained attention increasingly due to the development of relating technologies. At the same time, data breaches happen every day. Such security threats of personal data consistently challenge trust relationships between different stakeholders in data management. For example, users expect the parties who collect and process their personal data being trustworthy and sufficient to protect their privacy, and the companies rely on these data from users to make decisions or use it for other purpose to benefit users. On the other side, the trustworthiness and integrity of data, which could affect the quality of data, have been doubted and furthery impact the decision of data utilization [9]. The mistrust between stakeholders in data managements results the overall untrustworthy personal data ecosystem. Furthermore, while the world has become ever more data-driven, the balance between data utilization and privacy protection has also become a concern. ITU-T has published the recommendation Y.3055 to overcome the untrustworthy personal data ecosystem by proposing a trust-based personal data management framework (TPDM). In this recommendation, stakeholders in TPDM are categorized to personal data principles, personal data controller, personal data processor and third parties, and the phases of personal data flow are defined as personal data management phase, data collection phase and data management phase. The framework contains the architecture and requirements for each function inside. The mechanism of trust provisioning proposed in ITU-T Y.3052 is suggested to apply to scenarios which requires trade-off between data utilization and privacy protection to enhance trust between stakeholders in data management, so that trustworthy personal data ecosystem could be achieved.

Another important issue related to trusted data is the isolated data island problem. The increased application of machine learning rises isolated data island problem which data set could not be combined due to regulation, competition or ethical considerations. IEEE 2830-2021, published by IEEE Standard Association, introduces a framework of trusted execution environment (TEE) based shared machine learning (SML) to resolve difficulties of large scale, multi-source data sharing and analysis as well as multiple participants collaboration authorization in machine learning model training. This publication constructs a verifiable basic framework which contains architecture, functional components and processing procedure, together with some scenarios to illustrate the application of such framework. It also outlines both technical requirements and security requirements for such SML to satisfy trust and security. Technical requirements are organized by categories that include basic requirements, scalability, reliability, compatibility, performance and usability. Security requirements are instead divided into four clusters: authentication, access control, security auditing and data security.

Isolated data also exist due to the regulation and policies of data in cloud computing where industrial cloud for common purposes is taking shape. ISO/IEC TR 23186:2018 presents a trust framework for the cloud processing of multi-sourced data to mitigating trust issues between cloud service providers (CSP), cloud service customers (CSC) and cloud service users (CSU) in the processing of multi-sourced data. It demonstrates the application of trust in occasions, such as transportation and automation where trusted processing of multi-sourced data is crucial, and evaluates the importance of trust in these occasions. The presented trust framework contains data use obligations and controls, data provenance, chain of custody, security and immutable proof of compliance as elements of trust with their usage in agreements, and provides a data flow for trusted processing of multi-source data.

*C. Trusted Algorithm*

Every digital transaction relies on the algorithm behind it to run. Thus, trusted algorithm is the key to the trustworthiness of the future digital world. Nowadays, most of these algorithms are AI-based. However, users could not trust the decision that AI makes since they do not understand how AI makes the decision. This is due to the fact that AI is more like a black box. The transparency, explainability, accuracy, and reliability are the challenges which trusted algorithms are facing [10].

Current standards mainly regulate AI from both user's perspective and AI's perspective considering AI trustworthiness. ISO/IEC TR 24028:2020 provides an overview of trustworthiness in AI. The main objective is to discuss possible approaches to mitigating vulnerabilities and challenges as well as improve trustworthiness of AI systems to identifying specific standardization gaps in relating field. By surveying current threats and risks to AI systems which may affects overall trustworthiness, ISO/IEC JTC 1/SC 42 suggests to establish trust through transparency, explainability, controllability and etc., and advices on trustworthiness assessments referring characteristics of trustworthy AI. The characteristics of AI in this standard are defined as availability, resiliency, reliability, accuracy, safety, security and privacy.

NISTIR 8332 instead concentrates on the user trust in AI. With the analysis of trust challenges in AI systems, it introduces an approach to calculate AI user trust. The calculation involves pertinence and sufficiency of AI trustworthy characteristics as well as user experience in AI systems. The ranking of each characteristic may be different depending on the occasions of AI systems. Comparing to the characteristics that ISO/IEC JTC 1/SC 42 suggests, NIST believes that accountability, objectivity and explainailtiy are also important, while availability is less important than others. Usability are measured by efficiency, effectiveness, and user satisfaction.

IV. TYPICAL TRUST APPLICATION SCENARIOS

At the top is trust application layer, and many standard organizations have published a series of standards to guide and regulate the application of trust. In this section, we present four typical trust application scenarios covered in existing standard.

*A. Trustworthy Network*

Traditional network security model, assumes that every entity inside the network are trusted. Entities from outside are untrusted and need to be authorized to join the network. Castle-and-moat model draws a clear network boundary between trusted zone and untrusted zone by allocating much more resources on defending external threats. Such approach is efficient to against external attack, but vulnerable to internal attack since an adversary who has gained the access to the network is considered as trusted entities in this case [11]. Therefore, ITU SG 13 proposed a framework of trustworthy networking with trust-centric network domains in ITU-T Y.3053, aiming to introduce trust provisioning in constructing trustworthy network. This recommendation describes a conceptual model of trustworthy networking which involves identification, trust evaluation and trustworthy communication. Entities inside the network domain rely on identification and trust evaluation to authenticate the entities they interact with, and perform trustworthy communication afterwards. The trust relationships for entities within the same network domain and with external entities are different as well. The document also provides integrated architecture and requirements for the actual deployment of framework and conceptual model.

Instead of introduce trust provisioning, NIST introduces the idea of zero trust and proposed a framework of zero trust architecture (ZTA) deployment in enterprise environment in NIST SP 800-207. Zero trust is defined as a term which assumes there is no implicit trust granted to elements in the network based on the physical location or network location or based on ownership. The boundary of network is blurred since the threats are assumed to be from both inside and outside of network, and thus strict verification before accessing to network or resources is required for every entity. Zero trust architecture gives planning of industrial and enterprise infrastructure and workflows based on zero trust principles. NIST provides the requirements to regulate ZTA deployment in enterprise environment as well as elaborations of possible interactions between ZTA and other federal guidance.

Another Standard closely related to trust networking is ITU-T X.1812. In this recommendation, X.1812 described application scenarios of 5G systems, and analyzed stakeholders and their trust relationships for each scenario. 5G has introduced more features and more open access environment than 4G, and thus involving more stakeholders in the ecosystem and complicates the trust relationships between different stakeholders. These new features such as virtualization and slicing, together with open access environment, making the network and service deployment more flexible, but also create vulnerability and making vague of the boundary between network and service in 5G. Thus, X.1812 proposed a security framework supported by the trust model. The trust model is designed based on trust relationship mapping, and the trust level and the trust criteria for the trust model are also clarified in the recommendation.

*B. Trustworthy IoT*

Compare to conventional devices, many IoT devices, such as sensors, have different ways interacting with the physical world. They are managed in a different way and are fragile under attack or malicious actions due to the limitation of hardware or architecture [12]. Current common security approach tends to create a closed network which only allows the devices from the same manufactures to join. Such approach could ensure that the devices in the IoT are trusted based on the trust of manufactures and simplify IoT environment, yet it against the idea of everything connects together. To solve this issue, trustworthy IoT is proposed as a promising solution.

The trust framework associated to IoT system and services is proposed by ISO/IEC JTC 1/SC 41 in ISO/IEC 30147:2021. This document focuses on the challenges in IoT systems that are not covered previously. The objective is to achieve trustworthy IoT systems by introducing system life cycle processes in the implementation and maintenance of trustworthiness in IoT systems. One of the significant features of IoT system is that it could be system of system (SoS), and it collaborates with other organization to operate and manage the constituent systems of the IoT system, which requires every system in IoT system to be trustworthy in order to achieve trustworthy IoT system. The characteristics of trustworthiness, including security, reliability, safety, privacy and resilience as well as risk of each characteristic in IoT systems are specifies. On top of ISO/IEC/IEEE 15288:2015, the document carries out a refinement and customization for the implementation of system life cycle processes in IoT systems, so that IoT systems which applies such system life cycle processes could achieve trustworthiness from dimensions of above characteristics.

As an important use case of IoT system, Intelligent Transport System (ITS) faces the trade-off between user data utilization and privacy protection likewise. Therefore, in ETSI TS 102 941 V2.1.1, TC ITS presents a framework of trust and privacy management in ITS communication to enhance security as well as build trust and security in ITS environment. It summarizes the required trust establishment and privacy managements for supporting security ITS environment, and clarifies the existed relationships between entities and elements of ITS reference architecture. It also lists required security services for privacy management in ITS. The included are ITS lifecycle management, public key infrastructure (PKI) and trust provision. For each security service, the document classifies the considerations, requirement and implementation details in actual deployment scenarios.

*C. Trustworthy Cloud*

As one of the service delivery models, Infrastructure as a Service (IaaS) greatly conveniences the provisioning and management by abstracting the hardware and allowing users to purchase server, network, storage and etc. as a service without worrying the complexities of deployment [13]. However, security and privacy of workloads has been a concern in current multi-tenant cloud environment. Each workload needs to be isolated to avoids mutually interference and access. Also, the migration of workloads between different cloud server is sometimes restricted by local relevant policies and laws, which demands trusted geolocation to determine the restriction of cloud server.

Therefore, the solution which combines hardware root of trust and trusted compute pool is proposed by NIST to realize trusted geolocation while deploying and migrating workloads between different cloud servers within a cloud. NIST suggested the organization to implement an automated hardware root of trust, together with the host's unique identifier and platform metadata in the hardware of cloud server to access geolocation information and enforce and monitor geolocation restriction. Such approach could guarantee the integrity of geolocation information and platform with the assumption of tamper-resistant hardware and firmware. Besides, trusted compute pool is required to achieve different workloads isolation by aggregating trusted systems and separating them from untrusted resources. The proof of concept implementation of the solution is proposed in NISTIR 7904. Based on this solution, National Cybersecurity Center of Excellence (NCCoE) develops NIST SP 1800-19 to describe the approach, architecture and security characteristics of this solution in details with evaluation of how such solution could provide the necessary security capabilities, and provides a sample solution with deployment details and prototype. NISTIR 8320A and NISTIR 8320B elaborate how the solution of trusted compute pool leveraging hardware root of trust with workload orchestration could be implemented to protect application container deployments in multi-tenant cloud environments instead. Workload orchestration could ensure that containers can only be instantiated on server platforms from satisfactory location which meet trustworthiness. Issue of decryption keys and initial encryption of container images may also be involved in orchestration.

ITU-T Y.3514 published by ITU SG 13 specifies the required security mechanism and overall trust framework to support establishment of trusted inter-cloud relationship among multiple cloud service providers (CSPs). Inter cloud, or "cloud of clouds" is the concept of interconnected multiple clouds which addresses the issue of limited resources in single cloud. Such concept allows CSPs corporate with one or more CSPs with relationship pattern of peering, federation or intermediary to maximize utilization of cloud resources. Interoperability and portability are highlighted for CSPs. Trusted relationships between CSPs and cloud service consumers (CSCs) or within multiple CSPs are essential to achieve trusted inter cloud computing successfully. Also, different security levels shall be considered in the management of trusted inter-cloud depends on the technology that CSCs and CSPs deploy. In Y.3514, the necessities and properties of trusted inter-cloud relationships are specified. The requirements according to the characteristics of governance, management, resiliency, security and confidentiality of trusted inter-cloud computing are included.

Isolation issues and confidentiality issues are the main security threats in inter-cloud systems. The potential problem from CSP's perspective is malicious user who threats to virtualization layer, isolation, server and so on, and the potential problem from CSC's perspective is data security and privacy. On top of ITU-T Y.3514, ITU-T SG13 expands the management framework of trusted inter-cloud computing and provides an overview of trust management in an inter-cloud environment in ITU-T Y.3517 to mitigate risk from threats mentioned previously. This framework involves isolation and security management mechanism, which is based on distributed cloud management, and enumerates scenarios for the implementation of such mechanism. ITU SG 13 also interprets an inter-cloud trust management model, two approaches for reputation-based trust management within inter-cloud environments, and a cloud service evaluation framework for inter-cloud trust management solution.

*D. Trustworthy Media*

Modern media environment has been changed and brings various content sharing methods nowadays with the development of ICT. In the past, the media environment is more like broadcasting, while user participates as receiver and media service providers (i.e. broadcast and mass media) are participates as senders, which makes content sharing restrict to user, but senders are mostly trusted and reliable by the mass [14]. Today, on the platforms such as Youtube and Tiktok, users participate as both senders and receivers, sharing and receiving information and content that are available. However, such freedom makes the environment highly complicated and risky. It is hard for users to evaluate whether the users interact with them are trustworthy or not. Adversary could be user as well and act maliciously. Such situation not only affect relationship between sender and receiver, but also impact trust relationship between service provider and service consumer(user).

ITU SG 13 identifies the potential risks in three categories: threats of media service, threats of content and threats of user privacy in ITU-T Y.3054. Current media service providers are not capable to against these risks and create trust and safe content sharing environment, since most of them rely on limited rating and comment mechanism. With such analysis ITU-T SG 13 proposed a framework for trust-based media services to overcome these limitations. The objective is to identify and mitigate the potential risks by stopping potential adversary to perform malicious action, which requires predictability and reliability from media service provider. Such framework allows media service providers to evaluate and utilize user trust by collecting, analyzing and modeling user data with trust management and trust model.

## VI. Future Work and Conclusions

The persistent working on standardization of trust has provided the basis for the shape of future digital world, yet there are more gaps anticipating resolution, and implementation is also required. It is clear that most of current trust related standardization works are focusing on network and computing more than other fields from previous summarization, but the specific fields which each standard focus are discrete. According to the information from ITU and IETF, ITU current working programs are still mostly focusing on trust in network, while working programs from IETF are more about TEE and related protocols. However, there are still numerous standardization gaps awaiting to be filled. For instance, present in forced standard in media only provides a solution to mitigate risks in the trust relationship between different stakeholders in the media environment. The content on media platform is also an important factor impacting the trust of service providers and content providers, and standardization is needed for evaluation of content trustworthiness. Moreover, the popularity of trust modelling has encouraged proposals of trust models implementing diversely. The performance of these trust models may be impacted by occasions, and the evaluation for the quality of trust model is required, which necessitates a general metric to standardize these trust models.

To conclude, this paper has provided a comprehensive overview of the state of the art in trust standardization by summarizing trust related standards grouped by fields and evaluating the key problems they resolved. Suggestion and discussion are made as well beyond the overview.

## Biographies

**Huilin Wang** (hwang56@u.nus.edu) is currently pursuing the B.Comp. degree in information security with National University of Singapore. The article is accomplished during her internship which mainly focuses on trust and trust modeling in communication network at the Digital Identity and Trustworthiness Laboratory, Huawei Singapore. Her research interests include digital security and network security and plans to pursue further studies in the areas.

**Xin Kang** (kang.xin@huawei.com) is senior researcher at Huawei Singapore Research Center. Dr. Kang received his Ph.D. Degree from National University of Singapore. He has more than 15 years' research experience in wireless communication and network security. He is the key contributor to Huawei's white paper series on 5G security. He has published 70+ IEEE top journal and conference papers, and received the Best Paper Award from IEEE ICC 2017, and Best 50 Papers Award from IEEE GlobeCom 2014. He has also filed 60+ patents on security protocol designs, and contributed 30+ technical proposals to 3GPP SA3. He is also the initiator and chief editor for ITU-T standard X.1365, X.1353, and the on-going work item Y.atem-tn.

**Tieyan Li** (Li.Tieyan@huawei.com) is currently leading Digital Trust research, on building the trust infrastructure for future digital world, and previously on mobile security, IoT security, and AI security at Shield Lab., Singapore Research Center, Huawei Technologies. Dr. Li is also the director of Trustworthy AI C-TMG and the vice-chairman of ETSI ISG SAI. Dr. Li received his Ph.D. Degree in Computer Science from National University of Singapore. He has more than 20 years experiences and is proficient in security design, architect, innovation and practical development. He was also active in academic security fields with tens of publications and patents. Dr. Li has served as the PC members for many security conferences, and is an influential speaker in industrial security forums.

**Zhongding Lei** (lei.zhongding@huawei.com) is a senior researcher at Huawei Singapore Research Center. He has been working on 5G network security since 2016. Prior to joining Huawei, he was a laboratory head and senior scientist with the Agency for Science, Technology, and Research (A-STAR) of Singapore, involved in research and development of 3GPP and IEEE standards in wireless systems and networks. He has been the Editor-in-Chief of IEEE Communications Standards Magazine since 2019.

**Cheng-Kang Chu** (Chu.Cheng.Kang@huawei.com) received his Ph.D. in Computer Science from National Chiao Tung University, Taiwan. He is a senior researcher of Huawei International, Singapore. Dr. Chu has had a long-term interest in the development of new technologies in applied cryptography, cloud computing security and IoT security. His research now focuses on mobile security, IoT security, decentralized digital identity, Web 3.0, etc. Dr. Chu has published many research papers in major conferences and journals like PKC, CT-RSA, AsiaCCS, IEEE TPDS, IEEE TIFS, etc. and received the best student paper award in ISC 2007.

**Haiguang Wang** (wang.haiguang.shieldlab@huawei.com) is a senior researcher on identity, trust and network security in Huawei International Pte. Ltd. He received a Ph.D. degree in Computer Engineering from National University of Singapore in 2009 and a Bachelor from Peking University in 1996. He joined Huawei at year 2013 and currently he is a senior researcher at Huawei. He was a research engineer/scientist at I2R Singapore since 2001. He is an IEEE Senior member.